\documentclass[usenatbib]{mnras}
\usepackage{graphicx}
\usepackage{amsmath}
\usepackage{txfonts,times}

\usepackage{booktabs}

\usepackage{xcolor}

\title[Bars at high redshift: Euclid TNG50 mocks]{The detectability of bars at high redshift: a case study using Euclid-like mock observations of TNG50 simulated galaxies}
\author[G. F. Gon\c{c}alves et al.]{Gustavo F. Gon\c{c}alves$^{1}$\thanks{E-mail:goncalvesg@alunos.utfpr.edu.br},
Rubens E. G. Machado$^{1,2}$,
R. R. Valen\c{c}a$^{2,3}$,
E. Athanassoula$^{4}$,\newauthor
Kar\'{\i}n Men\'{e}ndez-Delmestre$^{5}$,
Thiago Bueno-Dalpiaz$^{6}$
\\
$^{1}$Departamento Acad\^emico de F\'isica, Universidade Tecnol\'ogica Federal do Paran\'a,
Rua Sete de Setembro 3165, Curitiba, PR, Brazil\\
$^{2}$Instituto de Astronomia, Geof\'isica e Ci\^encias Atmosf\'ericas, Universidade de S\~ao Paulo, Rua do Mat\~ao 1226, S\~ao Paulo, SP, Brazil\\
$^{3}$European Southern Observatory, Karl-Schwarzschild-Strasse 2, 85748 Garching bei M\"unchen, Germany\\
$^{4}$Aix Marseille Universit\'e, CNRS, CNES, LAM, Marseille, France\\
$^{5}$Universidade Federal do Rio de Janeiro, Observat\'orio do Valongo,
Ladeira Pedro Ant\^onio 43, Sa\'ude, CEP 20080--090, Rio de Janeiro, RJ, Brazil\\
$^{6}$Centro Brasileiro de Pesquisas F\'isicas, Rio de Janeiro, RJ, 22290--180, Brazil
}

\date{Accepted XXX. Received YYY; in original form ZZZ}
\pubyear{2025}

\begin{document}  
\label{firstpage}
\pagerange{\pageref{firstpage}--\pageref{lastpage}}
\maketitle

\begin{abstract}
% context
Modern surveys such as Euclid report a decline in the fraction of barred galaxies from the local Universe to $z \sim 1$, whereas the TNG50 simulation predicts higher bar fractions, in tension with observations. This discrepancy may be due to observational biases in bar detectability when comparing simulations with observations.
% aims
We present a proof-of-concept study quantifying how Euclid-like observational conditions affect bar detectability in TNG50.
% methods
We analysed the entire galaxy sample at $z=0.5$ and highlight one borderline case with a bar length of 2.1\,kpc and strength $A_2=0.4$. Synthetic images were produced with Monte Carlo radiative transfer and realistic post-processing, and analysed with ellipse fitting and Fourier decomposition, as well as the recently constructed Zoobot analysis. Results were compared to idealised, noise-free stellar mass maps.
% results
In the illustrative case the bar is clearly detected in the mass map and remains visible in the Euclid VIS $I_{\rm E}$ filter, where Zoobot also classifies it as barred, but becomes undetectable in $Y_{\rm E}$ and in the VIS--NISP RGB composite, with all methods failing outside VIS. Extending to the full $z=0.5$ sample, Zoobot recovers only 31/141 galaxies, while $A_2$ and ellipse fitting perform better (80/141 and 67/141) but still miss many short or weak bars. When non-detections are counted as unbarred, the bar fraction of 44 per cent falls to \mbox{12--33} per cent depending on method. These results demonstrate the strong impact of observational effects on bar detectability and motivate bar-fraction estimates which incorporate realistic instrumental conditions across redshift in cosmological simulations.
\end{abstract}


\section{Introduction}

Stellar bars are common features in spiral galaxies in the local Universe, playing a key role in the secular evolution of galaxies. The observed bar fraction is known to evolve with redshift, reflecting both the intrinsic evolution of bars over cosmic time and the increasing observational challenges in detecting them at higher distances.

Progress in this field relies on the combined use of large observational surveys and cosmological simulations, which together provide complementary insights into the mechanisms governing bar formation and evolution. Understanding how the bar fraction evolves with redshift is essential for constraining models of galaxy formation and secular evolution. This is closely linked to the dynamical influence of bars on different galactic components, including the disk, bulge, and halo, primarily through angular momentum redistribution \citep{Athanassoula2003}.

Quantifying the observed bar fraction as a function of redshift is a delicate question in its own right, due to the different selection criteria that lead to considerable spread among different observational results. Additionally, comparisons between observations and simulations are even more fraught with uncertainty. Even at low redshift, the exact value of the local bar fraction depends on the classification methodology, instrumental characteristics and sample selection criteria---particularly regarding the inclusion or exclusion of weak bars. For all these reasons, the local bar fraction can range from 25 to as much as 70 per cent \citep[e.g.][]{Barazza2008, Aguerri2009, Masters2011, Cheung2013, Diaz-Garcia2016, Erwin2018}. Nevertheless, it is found observationally that the fraction of bars decreases with increasing redshift \citep[e.g.][]{Sheth2008, Melvin2014}. Systematic differences also arise from the visual classification methodology itself. Studies based on Galaxy Zoo volunteer classifications typically report lower local bar fractions \citep[around 30 per cent;][]{Masters2011}, whereas visual classifications conducted by professional astronomers often yield significantly higher values, frequently exceeding 60 per cent \citep[e.g.][]{Eskridge2000}. This discrepancy is commonly associated with differences in the interpretation and classification of weak bars by citizen science volunteers compared to professional astronomers.

Several observational and methodological factors have been reported in the literature as key challenges for reliably identifying bars in galaxies. Spatial resolution is a critical limitation, especially at high redshifts, where small-scale features become harder to resolve \citep{Menendez-Delmestre2007}. The choice of wavelength and filter also plays an important role \citep{Menendez-Delmestre2024}, as bars are more easily detected in near-infrared bands, which better trace the old stellar population and are less affected by dust obscuration \citep{Eskridge2000}. Bar size are also important factors affecting detectability. Short bars are intrinsically more difficult to identify and characterize, especially at higher redshifts where their typical angular size becomes smaller and increasingly comparable to the instrumental resolution \citep{Sheth2003}. Thus, many short bars may fall below the angular resolution limit of the instrument. If the population of such short bars is appreciable at high redshifts, this would significantly impact the interpretation of the bar fraction evolution.

Different approaches have been employed to identify bars in galaxies, each with its own strengths and limitations. Traditional visual classification has a long history in extragalactic astronomy, dating back to early works such as \cite{Vaucouleurs1963} and, more recently, large citizen science projects like Galaxy Zoo \cite[e.g.][]{Geron2025, Simmons2014}. Quantitative methods based on two-dimensional photometric decomposition are also widely used, relying on tools such as IMFIT \citep{Erwin2015}, BUDDA \citep{2004deSouza}, and GALFIT \citep{Peng2002} to separate the contributions of bulge, disk, and bar components. Other techniques include isophotal ellipse fitting \citep{Jedrzejewski1987}, which identifies bar features through characteristic radial variations in ellipticity and position angle, and Fourier decomposition, where the strength of the $m=2$ mode ($A_{2}$) is often used as a quantitative bar indicator \cite[e.g.][]{Buta2006, Diaz-Garcia2016}. More recently, unsupervised machine learning techniques have also been explored as alternative methods for bar identification in large datasets, as we will discuss in the context of the first Euclid data release Q1 \citep{EuclidCollab2025} and its initial bar fraction analysis.

The Euclid Space Telescope Q1 represents a major advance in our ability to study galaxy morphology from local galaxies at $z=0.1$ to intermediate-redshift galaxies at around $z=1.0$, thanks to its wide-area coverage and unprecedented spatial resolution in the optical and near-infrared. The first public data release \citep[Q1;][]{EuclidCollab2025} has already enabled the measurement of the bar fraction in a sample roughly an order of magnitude larger than previous studies covering a similar redshift range, such as those based on the Hubble Space Telescope within the Galaxy Zoo framework \citep{Melvin2014}. In this context, the use of unsupervised machine learning models has become essential to efficiently classify the vast number of galaxies expected from future Euclid observations. A example is the Zoobot model \citep{Walmsley2023}, which was trained on volunteer classifications from the Galaxy Zoo project, based on Euclid galaxy images. During the volunteer classification phase, participants viewed both RGB composites---generated from the VIS and NISP filter channels---and single-channel VIS images, the latter providing higher spatial resolution \citep{Walmsley2025}. These human-labeled datasets were then used to train the Zoobot, enabling automated classification of the remaining thousands of galaxies in the Q1 release.

Cosmological simulations have been widely used to study the evolution of the bar fraction, offering theoretical baselines for comparison with observations. While TNG50 \citep{Nelson2019}, as shown by the bar-fraction analyses of \citet{Rosas-Guevara2022} and \citet{2022Zana}, stands out for reproducing a local ($z=0$) bar fraction that closely matches observational estimates, earlier simulations such as Illustris \citep{Peschken2019}, EAGLE \citep{Algorry2017}, and New Horizon \citep{Reddish2022} reported significantly lower fractions or even an absence of barred galaxies at low redshift. The Auriga simulation \citep{Fragkoudi2025}, though limited to a zoom-in sample of Milky Way analogs, shows improved agreement at $z=0$ and a more realistic redshift evolution when compared to previous works. Despite these advances, it is important to note that bar measurements in simulations are typically performed under idealised conditions, without accounting for observational limitations such as resolution, noise, or surface brightness dimming that affect real data. This discrepancy becomes especially relevant when comparing simulated bar fractions at higher redshifts.

\citet{Erwin2018}, working with Spitzer and SDSS data, has pointed out that discrepancies may be explained by the underestimation of bar fractions due to poor spatial resolution; and that even the resolution of HST is not enough to prevent small bars from being missed at high redshift. More recently, initial measurements of bars from JWST data have been performed at considerably high redshifts. \citet{Guo2025} find a decreasing fraction of bars reaching $\lesssim$10 per cent at $z\sim2-4$, but point out that TNG50 predicts a large population of smaller bars that the JWST data could not be detect. \citet{2026LeConte} further show, using JWST/NIRCam data at $1 \leq z \leq 4$, that although the bar fraction declines with redshift, bars can already be long and strong at $z>1$, with bar and disc sizes evolving in tandem. \citet{Liang2024} performs a interesting study of image degradation by artificially shifting one nearby observed spiral into progressively larger redshifts and producing simulated images comparable to JWST ones at those redshifts. They find that galaxy remains detectable as long as its bar length is greater than twice the full width at half maximum (FWHM) of the instrument PSF used (JWST in \citealt{Liang2024}), similar to a factor of 2.5 (derived for SDSS FWHM) obtained by \citet{Aguerri2009} via extensive experiments on artificial galaxies. \citet{LeConte2024} recently analysed JWST galaxies in the redshift range $1 \leq z \leq 3$ and found a decrease from 17.8 to 13.8 per cent in the bar fraction. They also point out that under poorer spatial resolution, the ellipticity of bars tends to be seen as rounder, making the bar less distinguishable. The conclusion is that bars shorter than $\sim$2.5--3\,kpc are probably missed in those redshifts.

Following the bar-fraction analysis of \citet{Rosas-Guevara2022}, who focused on galaxies with stellar masses in the range $10^{10} \lesssim M_\star /{\rm M_\odot} \lesssim 10^{11}$, we examine how the evolutionary trend reported for TNG50 compares with observational expectations. Intriguingly, TNG50 exhibits an unexpected evolutionary trend: the bar fraction increases with redshift, showing a higher value at $z\sim1$ than at 
$z=0$, which seems inconsistent with observational results. An increase of bars from $z=0$ to $z=1$ is not seen in observations. This raises an important question: to what extent could observational biases---particularly those affecting the detectability of bars at high redshift---explain or mitigate this apparent discrepancy? Addressing this issue by quantifying the impact of observational effects on bar detectability measurements in simulations is the main goal of this work. Throughout this analysis, the simulated galaxies considered correspond exactly to the \citet{Rosas-Guevara2022} sample at $z=0.5$.

Another source for assessing the fraction of barred galaxies as a function of redshift in TNG50 is presented by \citet{2022Zana}, who analyse an expanded galaxy sample and explore the impact of different selection criteria using morphological decompositions performed with the \textsc{MORDOR} (MORphological DecOmposeR) code. \textsc{MORDOR} operates directly on the stellar particle distribution and in this framework, bars are identified via the $m=2$ Fourier component, which traces departures from axisymmetry, performing a two-dimensional decomposition. The paper shows that the inferred evolution of the bar fraction is sensitive to the adopted criteria. In particular, when the most massive galaxies are excluded from the sample, the previously reported increase in the bar fraction with redshift becomes approximately constant, highlighting the dependence of the measured trend on sample definition. Nevertheless, even under these alternative selection criteria, the excess of barred galaxies at high redshift reported in TNG50 remains present in their analysis.

Compared to \citet{Rosas-Guevara2022}, the analysis of \citet{2022Zana} is based on a larger sample, particularly enriched in low-mass galaxies ($M_\star < 10^{9.5}~{\rm M_\odot}$) and systems hosting small and weak bars. For galaxies at higher stellar masses, however, the galaxies IDs included in both studies are effectively identical. Accordingly, within the stellar-mass interval analysed by \citet{Rosas-Guevara2022}, the evolution of the bar fraction shows no substantial differences between the two analyses. For these reasons, and in line with the considerations above, as well as the minimum particle-resolution requirements for the construction of reliable mock observations, our sample is restricted to galaxies with $M_\star \geq 10^{10}~{\rm M_\odot}$ throughout this work. Motivated by these considerations, the list of galaxy IDs considered in our analysis are taken from \citet{Rosas-Guevara2022}.

Throughout the paper, a standard cosmology is adopted with the same parameters as the IllustrisTNG simulations \citep{Nelson2019}, consistent with the \citet{Planck2016} results: $\Omega_\Lambda=0.6911$, $\Omega_{\rm m}=0.3089$ and $h=0.6774$. This paper is focused on $z=0.5$, which corresponds to a lookback time of approximately 5.19\,Gyr. At that redshift, 1\,kpc subtends approximately 0.159\,arcsec.

This paper is structured as follows. In Section~\ref{methods}, we describe the simulation used, the selection of the target subhalo, and the methods employed to create realistic mock observations, including radiative transfer (RT) modeling. In Section~\ref{results}, we present our results for bar detection, comparing the idealised, noise-free case with the realistic mocks incorporating observational effects. In Section~\ref{discussion}, we discuss the broader implications of this proof-of-concept study for interpreting bar fractions in cosmological simulations. Finally, Section~\ref{conclusions} summarises our conclusions.

\begin{figure*}
\centering
\includegraphics{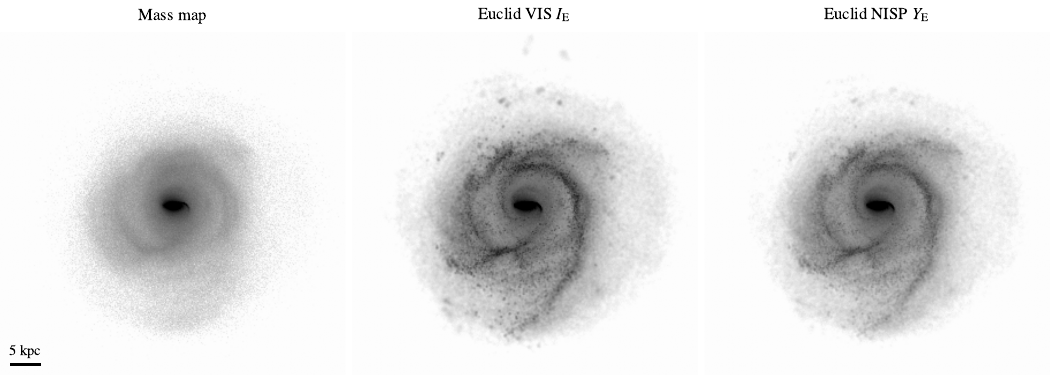}
\caption{Comparison of the stellar mass map and noise-free mock Euclid images for the barred galaxy ID184179. From left to right: the stellar mass map, the synthetic Euclid VIS $I_{E}$ image and the synthetic Euclid NISP $Y_{E}$ image, each covering a physical size of 60 × 60 kpc and shown in a face-on orientation. No observational noise or PSF convolution was added.}
\label{fig:fig1}
\end{figure*}

\section{Mock images}
\label{methods}

In this section, we will present the methods used to produce Euclid mock images from TNG50 galaxies. A visualization is already shown in Fig.~\ref{fig:fig1}, which gives the maps of stellar mass, alongside mock images in the two relevant Euclid filters, which will be explained in what follows. These are noise-free view, and thus serve to highlight only the contrast between using mass and light (at different wavelengths) to view a simulated galaxy.

\subsection{The Illustris TNG50 simulation}

TNG50 is the highest-resolution run within the IllustrisTNG cosmological simulation suite \citep{Nelson2019,Pillepich2018}. Its spatial and mass resolution make it particularly suitable for studying sub-galactic structures such as stellar bars. The simulation follows the evolution of dark matter, gas, stars, and black holes within a cosmological box with a side-length of \(51.7 ~\mathrm{Mpc}\), using a total of \(2 \times 2160^3\) resolution elements. The dark matter particle mass is \(3.1 \times 10^5~\mathrm{M_{\odot}}\), and the mean baryonic mass resolution is \(8.5 \times 10^4~\mathrm{M_{\odot}}\). The minimum cell size for gas in the adaptive mesh is \(72~\mathrm{pc}\), while the gravitational softening length for dark matter and stellar particles is \(0.288~\mathrm{kpc}\) at \(z=0\) \citep{2019Pillepich}.

The galaxy used in this study was selected from the barred galaxy catalog of \citet{Rosas-Guevara2022}, which identified bars in TNG50 based on the bar strength obtained from the Fourier decomposition of the stellar mass distribution. Bar strength was defined as the peak of \(I_2\) value (see Eq.~\ref{eq:I2})\footnote{Rather than the $I_2$ notation adopted here, the same quantities are sometimes referred to as $A_2$ and $A_{2,\max}$ in the literature.}, while bar length was determined from the radial extent over which the phase of the \(m=2\) mode remained approximately constant. To minimize contamination from transient features, the catalog includes only galaxies with \(A_{2,\mathrm{max}} > 0.2\) and bar lengths exceeding a resolution-dependent minimum radius. Throughout this paper, the expressions bar length, bar radius or bar size always refer to the semi-major axis of the bar.

\begin{figure*}
\centering
\includegraphics{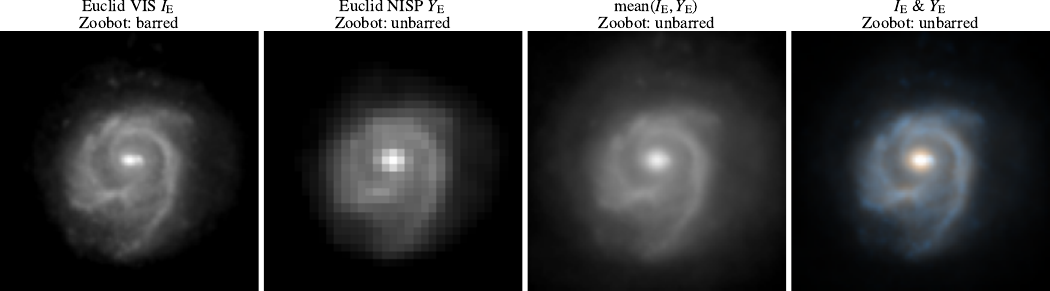}
\caption{Mock Euclid images of the barred galaxy ID184179 at \(z \sim 0.5\) in the $I_{E}$ and $Y_{E}$ bands, their pixel-wise mean, and an RGB composite image. The first three panels (monochromatic) show the $I_{E}$ band, the $Y_{E}$ band, and their mean image, each displayed using the monochromatic color scale adopted for Zoobot classification, after rebinning, synthetic noise addition, and PSF convolution. The fourth panel shows an RGB composite where the $Y_{E}$ band maps to the red channel, the $I_{E}$ band to blue, and the mean image to green. The Zoobot classification (barred or unbarred) for each image is shown at the top of each panel. All images span 60~$\times$~60\,kpc and are displayed face-on.}
\label{fig:fig3}
\end{figure*}

\begin{figure}
\centering
\includegraphics[width=\columnwidth]{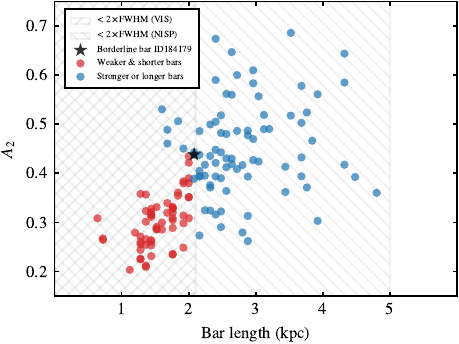}
\caption{Position of subhalo ID184179 within the bar parameter space at \(z \sim 0.5\), based on the distribution of bar radius (in kpc) and bar strengths (\(A_2\)) for all barred galaxies in snapshot~67 from the catalog of \citet{Rosas-Guevara2022}. Galaxies with both shorter and weaker bars than our target are shown in red, the remaining barred galaxies appear in blue, and ID184179 is highlighted in black. The hatched shaded regions along the \(x\)-axis mark the bar-length scales corresponding to \(2\times\mathrm{FWHM}\) of Euclid VIS (\(\sim 2.1\,\mathrm{kpc}\)) and NISP (\(\sim 5.0\,\mathrm{kpc}\)) at this redshift. This figure illustrates that the selected galaxy occupies an intermediate position in the parameter space, with many galaxies hosting even smaller and weaker bars at this redshift.} 
\label{fig:fig2}
\end{figure}

For this work, we selected subhalo ID184179 at snapshot~67, corresponding to \(z = 0.5\), which hosts a bar with a semi-major axis length of \(2.1~\mathrm{kpc}\) and a bar strength of \(A_{2,\mathrm{max}} = 0.44\), placing it near the threshold between weak and strong bars according to the catalog criteria as well as by Zoobot detectability (see Fig.~\ref{fig:fig3}), $A_{2}$, and ellipse fitting in the mock images. This galaxy was chosen as a case study because it occupies an intermediate position in the bar parameter space at this redshift (Fig.~\ref{fig:fig2}). Its relevance is reinforced by the fact that many galaxies in the same snapshot exhibit even shorter and weaker bars, making this borderline case a useful benchmark for exploring detectability limits. In addition, its bar length lies exactly at the \(2 \times \text{FWHM}\) limit of the VIS $I_{E}$ filter, the band with the best spatial resolution in this analysis, further emphasizing its role as a critical threshold case. This can be seen in the hatched region of Fig.~\ref{fig:fig2}, where the red points provide a representative approximation of the hypothesis that bars shorter than this limit are unlikely to be reliably detected.

The methods exemplified with this borderline galaxy were then applied to the entire $z=0.5$ sample, allowing us to assess bar detectability and bar fraction estimates in a statistically representative context.

\subsection{SKIRT radiative transfer code and synthetic noises}

The radiative transfer simulations for this work were performed using the SKIRT9 code \citep{Baes2020}, a three-dimensional Monte Carlo radiative transfer (RT) code designed to model dust attenuation and emission processes in galaxies. SKIRT self-consistently computes key physical processes, including scattering, absorption, and thermal dust emission \citep{Baes2015}. It features a dedicated import interface compatible with outputs from hydrodynamical simulations such as IllustrisTNG, allowing the use of stellar particles as radiative sources and deriving the spatial distribution and properties of dust from the gas particle data.

The configuration steps and parameter choices for our SKIRT runs closely follow the procedures described in \citet{Trvcka2022} and further applied in \citet{Baes2024}, both of which are optimized to balance computational efficiency with physical realism.

The primary radiation field is generated from the stellar particles in the simulation, divided into two distinct components based on stellar age. Older stellar populations (age \(>10~\mathrm{Myr}\)) are modeled using the \citet{Bruzual2003} spectral energy distribution (SED) library, which provides a detailed treatment of stellar evolution assuming a \citet{Chabrier2003} initial mass function (IMF). Younger stellar populations, corresponding to star-forming regions (\(\leq10~\mathrm{Myr}\)), are modeled using the MAPPINGS~III library \citep{Groves2008}, which accounts for the complex interplay between massive stars and their surrounding ionized gas.

The dust component is derived from the properties of the gas particles in the simulation. We adopt the THEMIS dust model \citep{Jones2017} to describe the interstellar dust properties. To separate the dense interstellar medium (ISM) from the hot circumgalactic medium (CGM), we apply the temperature-density threshold criterion proposed by \citet{Trvcka2022}, which builds upon the gas phase classification methodology described in \citet{Torrey2019}.

Following the radiative transfer stage, SKIRT generates noise-free FITS images of the galaxy. The convolution with the transmission curves of the relevant Euclid filters is performed using the Python Toolkit for SKIRT9 (PTS)\footnote{\url{https://skirt.ugent.be}}. Examples of these noise-free, filter-convolved images are shown in Fig.~\ref{fig:fig1}. These outputs then serve as input for subsequent post-processing steps, where we apply realistic instrumental effects, including synthetic noise addition, spatial rebinning to match the filters angular resolution at the target redshift, and point spread function (PSF) convolution.

To place our post-processing strategy in context, we briefly note previous papers that applied similar instrumental degradations to simulation-based mock images. The effect of applying post-processing steps to SKIRT outputs in order to mimic an astronomical instrument has been addressed by recent papers in contexts other than evaluating the impact on bar detectability. We can cite \citet{2025Gong}, who created SDSS mocks for the decomposition of morphological parameters, as well as \citet{2019Rodrigues-Gomez}, who constructed Pan-STARRS mocks to obtain non-parametric indices for comparison with real Pan-STARRS images, as well as \citet{2023Eisert}, who used TNG galaxies and SKIRT to emulate Subaru/HSC observations and applied a machine-learning-based morphological classifier. All these works applied the instrumental PSF and added synthetic noise to originally noise-free images. These precedents motivate our adoption of an analogous post-processing pipeline tailored to Euclid-like observational conditions.

To simulate realistic observational conditions, we post-processed the noise-free SKIRT outputs by applying instrumental effects corresponding to the Euclid VIS and NISP filters. First, each noise-free FITS image was resampled (regridded) to match the pixel scale of the respective Euclid filter at the target redshift (\(z=0.5\)). For the VIS $I_{E}$ channel, a pixel scale of \(0.1~\mathrm{arcsec/pixel}\) was adopted, while for the NISP channel $Y_{E}$, we used \(0.3~\mathrm{arcsec/pixel}\), reflecting the differences in spatial resolution between the instruments \citep{Cropper2025, Jahnke2025}.

We then convolved each image with a Gaussian PSF characterized by the FWHM of each instrument. For VIS, we adopted a FWHM of \(0.17~\mathrm{arcsec}\), while for NISP the FWHM was set to \(0.4~\mathrm{arcsec}\), following the approach used in \citet{Abdurro2025}. Synthetic noise was added in two steps: first, Poisson noise based on the expected photon counts from the SKIRT flux outputs and filter zero-point; second, Gaussian sky background noise, with surface brightness levels consistent with typical Euclid survey depths.

After these steps, the final simulated FITS images include realistic spatial resolution, instrumental PSF effects, and synthetic noise representative of expected Euclid survey conditions. These images were used as the basis for our detectability analysis presented in the following sections.

\subsection{Zoobot machine learning model}

Zoobot \citep{Walmsley2023} is a deep-learning-based framework for classifying galaxy morphology, trained on millions of volunteer responses from previous Galaxy Zoo projects. The framework is available through a dedicated Python package, with its source code openly distributed via GitHub\footnote{\url{https://github.com/mwalmsley/zoobot}}. Comprehensive tutorials are provided to guide users in accessing the pre-trained models\footnote{\url{https://zoobot.readthedocs.io/en/latest/pretrained_models.html}}, available for both multi-channel and grayscale input images, as well as in fine-tuning Zoobot for custom prediction tasks.

Zoobot has already been applied to Euclid data in various contexts, including the analyses of the Q1 public data release for bar fraction measurements \citep{Walmsley2025} and pre-release studies making morphological predictions on Euclid observations \citep{Aussel2024}. The publicly available Zoobot models are trained on all previously published Galaxy Zoo datasets, incorporating the extensive set of morphological labels provided by volunteers across multiple project versions, which gives them a broad and representative training base. The resulting model is directly suitable for application to our radiative transfer mock images with realistic noise and instrumental effects.

\section{Results}
\label{results}

In this section, we evaluate bar detectability using three methods: the Zoobot machine learning classifier, ellipse fitting, and Fourier decomposition. These methods are applied to the radiative-transfer mock images, where the bar signal is subject to observational limitations, and compared to the corresponding stellar mass maps, where the bar is always clearly visible. We begin with a chosen borderline galaxy to illustrate the differences between methods and filters, and then extend the same analysis to the entire $z=0.5$ sample in order to assess the impact of observational effects on bar classification and bar fraction estimates.

\begin{figure*}
\centering
\includegraphics{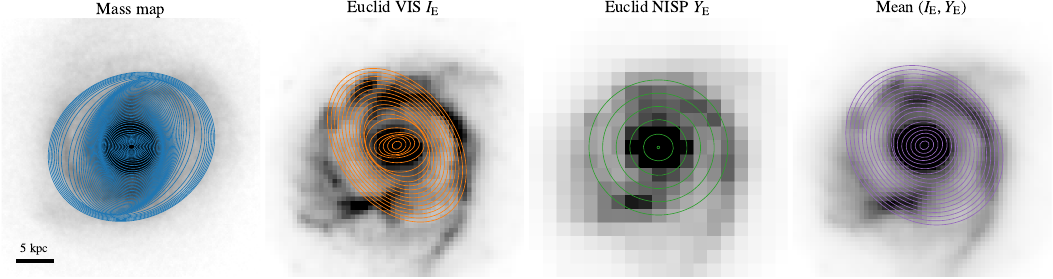}
\caption{Stellar mass map and mock Euclid images of the barred galaxy ID184179, with overplotted ellipse fits for each panel. From left to right: the stellar mass map, the mock Euclid $I_{E}$ image, the mock Euclid $Y_{E}$ image, and the pixel-wise mean of $I_{E}$ and $Y_{E}$, all shown face-on and covering 36 $\times$ 36 kpc. Ellipse fits derived from isophotal analysis are overplotted in blue (mass map), orange ($I_{E}$), green ($Y_{E}$), and purple (pixel-wise mean), respectively.}
\label{fig:fig4}
\end{figure*}

\subsection{Bar detectability via Zoobot classification}

As a proof-of-concept application, we tested the public Zoobot model on our mock observations of subhalo ID~184179, a TNG50 galaxy at \(z=0.5\) hosting a medium bar near the resolution limit of Euclid imaging. The galaxy's face-on orientation and intermediate bar properties make it a representative and challenging case for testing bar detectability under Euclid-like conditions. The classification tests demonstrate that the public model is effective in identifying bars in favourable conditions, especially for strong and extended bars.

In Fig.~\ref{fig:fig3} the following four mock images are shown. First, the image in the VIS $I_{E}$ band, which is the higher resolution instrument. Second, the image in the NISP $Y_{E}$ band, which has relatively lower resolution. Third, a monochromatic image which is a pixel-wise mean of $I_{E}$ and $Y_{E}$. Fourth, an RGB composite image, where the colors are attributed, respectively, as $Y_{E}$ (red channel), $(I_E+Y_E)/2$ (green channel), and $I_{E}$ (blue channel), as previously done in RGB compositions from the Euclid Collaboration \citep{EuclidCollab2025}. Because RGB composites are part of the visual material used in Euclid Galaxy Zoo classification, it is important to test whether they preserve bar signatures under Euclid-like conditions. In other words, the fourth frame of Fig.~\ref{fig:fig3} is the RGB composite of the previous three frames. The red channel receives the NISP image because it is the larger wavelength (near infrared), compared to VIS and green channel is the mean of the pixelwise flux in the other two channels. iven its role in Euclid visual classification, we also evaluate bar detectability in an RGB composite representation.

The four images in Fig.~\ref{fig:fig3} were fed to Zoobot, which returns class probabilities for the bar question: the probability that a bar is considered strong ($p_\text{bar,strong}$), the probability that a bar is considered weak ($p_\text{bar,weak}$), and the complementary probability of being non-barred ($p_\text{bar,no}$), such that the three add up to unity. In this work we define the bar probability as $P_\text{bar} \equiv p_\text{bar,strong}+p_\text{bar,weak}$, with galaxies classified as barred when $P_\text{bar} \ge 0.5$ and, consequently, as non-barred when $p_\text{bar,no} > 0.5$. For this case-study galaxy, Zoobot returned a barred classification only when using the monochromatic mock image from the Euclid $I_{E}$ filter. This result is illustrated in Fig.~\ref{fig:fig3}, which shows the Zoobot classification for each mock image variant. This outcome can be attributed to the higher spatial resolution provided by this filter, which enhances the visibility of small-scale structures such as bars. In contrast, the lower spatial resolution of the $Y_{E}$ filter from the NISP instrument clearly suppresses the detectability of the bar, making its non-detection unsurprising given the relative size of the structure compared to the instrument angular resolution.

Furthermore, in the RGB composite images created from VIS-NISP channels, where the spatial resolution of the VIS channel is known from previous studies \citep{EuclidCollab2025} to dominate the classification outcome, the averaging process across filters (pixel-wise mean) resulted in additional smoothing, an effect driven mainly by the larger pixel scale and the broader point spread function (PSF) associated with the $Y_{E}$ instrument. This smoothing effect was sufficient to suppress the detectability of the bar in subhalo ID184179, highlighting the sensitivity of bar detection to both resolution and image processing choices.

Thus, for this specific galaxy, the bar was detected only in the higher-resolution \(I_{E}\) mock image. In the \(Y_{E}\) filter and the VIS-NISP composite image, the model failed to identify the bar, likely due to resolution-driven smoothing effects.

\subsection{Bar detectability via ellipse fitting}

\begin{figure}
% \centering
\includegraphics[width=\columnwidth]{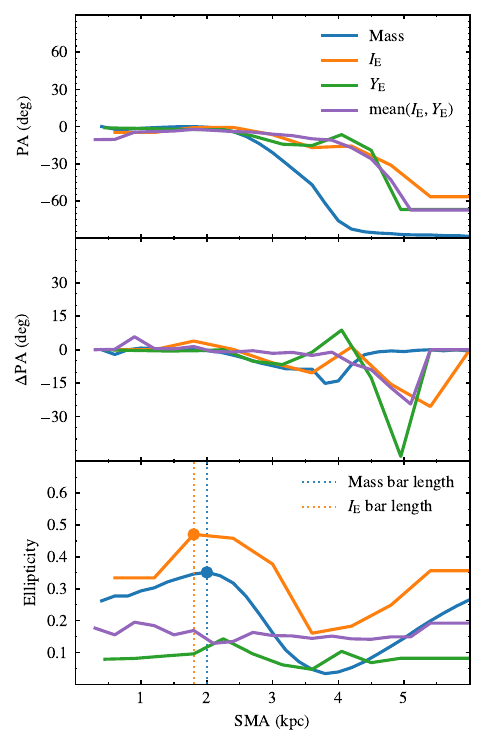}
\caption{Ellipse fitting results for the barred galaxy ID184179 in the stellar mass map and mock Euclid images. From top to bottom: position angle (PA), PA variation ($\Delta$PA), and ellipticity as functions of semi-major axis (SMA) in kpc. Continuous lines show measurements for each image: blue (mass map), orange ($I_{E}$), green ($Y_{E}$), and purple (pixel-wise mean of $I_{E}$ and $Y_{E}$). In both lower panels, points and dashed vertical lines mark the SMA of maximum ellipticity and the SMA where PA deviates from its inner constant plateau, both used as bar length estimates.}
\label{fig:fig5}
\end{figure}

We applied the isophotal ellipse fitting method to both the stellar mass map and the mock images to assess bar detectability (Fig.~\ref{fig:fig4}). As shown in Fig.~\ref{fig:fig5} in the stellar mass map, the bar is clearly visible, and the resulting ellipse fits show the expected behavior for a barred galaxy: a plateau in position angle (PA) across the bar region followed by a steep decline, and also a pronounced peak in ellipticity, both of which serve as standard indicators for bar length estimation. We measured its properties in this idealised dataset as a reference to evaluate how instrumental effects impact the bar observable signatures in the mocks.

Among the mock images, the bar meets the standard detectability criteria only in the $I_{E}$ filter. Across all filters, the PA remains approximately constant along the bar length and then changes abruptly near the bar end (first panel of Fig.~\ref{fig:fig5}), with this PA variation occurring at larger SMA (semi major axis) values in all filters compared to the mass map. Additionally, in the third panel, the ellipticity profile in the $I_{E}$ mock presents a local maximum at the bar radius, which can be used to estimate the bar length. In this case, the peak ellipticity is even higher in the VIS $I_{E}$ filter than in the stellar mass map, strengthening the bar signature rather than suppressing it. By contrast, in the other filters the PSF smoothing dominates, and no clear ellipticity maximum is visible, rendering the bar undetectable. While the middle panel suggests some bar-like signature in the PA profiles across different filters, the bottom panel of Fig.~\ref{fig:fig5} confirms that a distinct ellipticity peak is only recovered in the stellar mass map and in VIS $I_{E}$. These signatures are shown in the ellipticity and PA profiles in Fig.~\ref{fig:fig5}.

In the third panel of Fig.~\ref{fig:fig5}, the comparison between the blue and orange lines shows that the ellipticity peak is actually enhanced in the VIS $I_{E}$ image relative to the stellar mass map, rather than attenuated. This particular galaxy had an intrinsically well-defined profile in the original stellar mass, such that the signature was preserved in the orange line. However, one can speculate that in a galaxy with a slightly weaker bar, the corresponding attenuation might have been sufficient to erase the ellipticity drop, as occurred in the other filters shown.

In contrast, for both the $Y_{E}$ filter and the mean-combined image (mean of $I_{E}$ and $Y_{E}$), the bar is not detectable using ellipse fitting. The PA profile may appear to show some bar-like variation, but the absence of a corresponding ellipticity peak clearly demonstrates that the bar is not detectable. For the $Y_{E}$ filter, this result is expected since its angular resolution is comparable to the physical size of the bar, making the presence of detectable signatures unlikely. However, the lack of bar signatures in the mean-combined map highlights an important point: for bars that are not particularly long or visually prominent, the higher angular resolution of the $I_{E}$ filter alone is not sufficient to guarantee detectability. The additional smoothing introduced when combining filters, particularly through the pixel-wise mean, can be enough to suppress the signatures required for bar detection and measurement.

\begin{figure}
\centering
\includegraphics[width=\columnwidth]{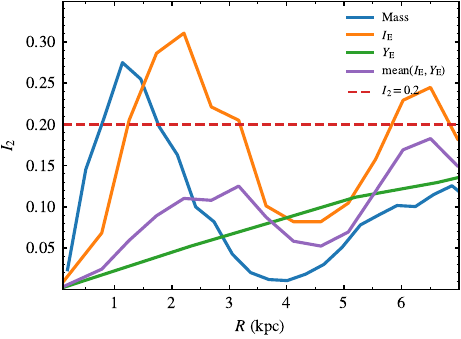}
\caption{Relative amplitude of the Fourier $m=2$ mode as a function of cylindrical radii in kiloparsecs (kpc) for the barred galaxy ID184179. Curves correspond to the stellar mass map (blue), $I_{E}$ (orange), $Y_{E}$ (green), and the pixel-wise mean of $I_{E}$ and $Y_{E}$ (purple). The horizontal red dashed line marks the threshold commonly used to identify significant non-axisymmetric structures.}
\label{fig:fig6}
\end{figure}

\subsection{Bar detectability via Fourier decomposition}

We also analyzed the stellar mass map and the mock images using Fourier decomposition to assess the detectability of non-axisymmetric structures, such as bars. The relative amplitude of the $m=2$ component of the Fourier decomposition of the projected mass (or light) is:
\begin{equation}
\label{eq:I2}
I_2 = \frac{\sqrt{a_2^2 + b_2^2}}{a_0} 
\end{equation}
and the coefficients are defined as:
\begin{eqnarray}
a_2 &=& \sum_i m_i \cos{2\theta_i}\\
b_2 &=& \sum_i m_i \sin{2\theta_i}\\
a_0 &=& \sum_i m_i
\end{eqnarray}
where $m_i$ is the mass and $\theta_i$ is the azimuth of each particle $i$. With this definition, the bar strength would be the peak value of $I_2(R)$, i.e.~$A_2=\max(I_2)$. Rather than the $I_2, A_2$ notation, sometimes the same quantities are named $A_2, A_{\rm 2,max}$ in other papers.

Applying a commonly used threshold for bar detection of $A_{2} > 0.2$ \citep[e.g.][]{2017Algorry, 2018Zana, 2021Fragkoudi, Rosas-Guevara2022}, only the idealised stellar mass map and the VIS $I_{E}$ mock would be classified as barred. In the VIS $I_{E}$ filter, the bar produces an $A_{2}$ peak that appears slightly stronger than in the mass map, consistent with the higher ellipticity values from ellipse fitting, although the maximum occurs at a somewhat larger radius. The blue peak in Fig.~\ref{fig:fig6} clearly confirms the presence of the bar in the mass map, as well as its known extent of close 2.1\,kpc according to the \cite{Rosas-Guevara2022} bar length definition. The orange line in Fig.~\ref{fig:fig6}, corresponding to VIS $I_{E}$, therefore shows a comparable $A_{2}$ profile with a peak shifted to larger radii. For the other mocks (green and purple lines in Fig.~\ref{fig:fig6}), no hint at all is present in the bar region. The bumps around 4--5\,kpc are merely capturing the mild non-axisymmetry of the spiral arms and the rest of the disc.

The similarities and differences between methods can be understood by the nature of the signatures each technique traces. While the PSF smoothing reduces the isophotal shape contrast in most mocks, the characteristic features used in ellipse fitting---such as the presence of a local ellipticity maximum and a nearly constant position angle (PA) along the bar region---are clearly recovered in the VIS $I_{E}$ filter and even appear slightly stronger than in the stellar mass map, although the peak occurs at a somewhat larger radius. In the other filters, however, these signatures vanish and the bar becomes undetectable. Fourier decomposition likewise recovers the bar only in VIS, with an $A_{2}$ profile comparable to the mass map but shifted to larger radii, while in the remaining filters the signal falls below the detection threshold and is lost in the surrounding noise.

\subsection{Sample-wide bar detectability at redshift 0.5}

We now extend the analysis beyond the illustrative borderline case and apply the same three methods---Zoobot, ellipse fitting, and Fourier $A_{2}$---to the entire $z=0.5$ sample using VIS-like mock images. Reference values for bar length and $A_2$ are required to characterise barred galaxies in the figures and comparisons, for this purpose, we adopt the noise-free measurements derived from stellar-mass distributions reported by \citet{Rosas-Guevara2022}, which are commonly used as a baseline.

\begin{figure}
\centering
\includegraphics[width=\columnwidth]{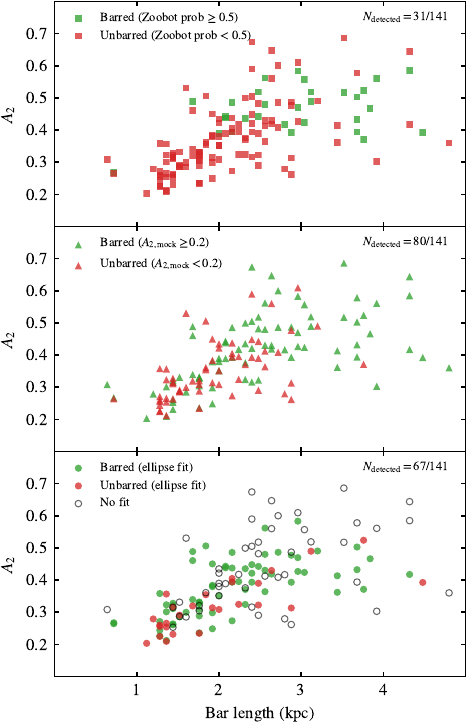}
\caption{Comparison of the detectability of barred galaxies in the bar radius–strength ($A_2$) plane space at \(z \sim 0.5\), based on galaxies classified as strong and weak barred galaxies in \citet{Rosas-Guevara2022}. Each panel shows the classification outcome from a different method applied to the same galaxy sample, with distinct symbols: top, Zoobot probability threshold ($p_{\mathrm{bar}} \geq 0.5$ = green squares, $p_{\mathrm{bar}} < 0.5$ = red squares); middle, Fourier amplitude criterion ($A_{2,\mathrm{max}} \geq 0.2$ = green triangles, $<0.2$ = red triangles); bottom, ellipse fitting (barred = green circles; unbarred = red circles; unfit = black open circles). All tests were performed on the VIS filter mocks, which proved to be the most suitable for bar detection.}
\label{fig:fig7}
\end{figure}

In Fig.~\ref{fig:fig7} we summarizes the method-by-method recoveries in the mass-map-based bar radius–strength plane from \citet{Rosas-Guevara2022}: Zoobot identifies 31/141 barred galaxies, the Fourier criterion recovers 79/141, and ellipse fitting recovers 66/141 (with an additional subset lacking reliable isophotal fits).

\begin{figure}
\centering
\includegraphics[width=\columnwidth]{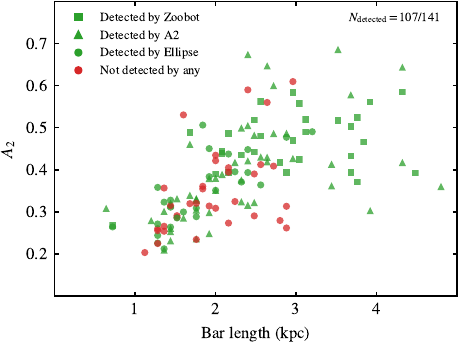}
\caption{Bar detectability in the bar radius–strength ($A_2$) plane for galaxies space at \(z \sim 0.5\) based on galaxies classified as strong and weak barred galaxies in \citet{Rosas-Guevara2022}. Each galaxy is represented by a single symbol according to the first method (in priority order: Zoobot, Fourier amplitude, ellipse fitting) that identifies it as barred: green squares (Zoobot, $p_{\mathrm{bar}}\geq 0.5$), green triangles (Fourier amplitude, $A_{2,\max}\geq 0.2$), and green circles (ellipse fitting). Galaxies not detected as barred by any method are shown as red circles.}
\label{fig:fig8}
\end{figure}

The single-galaxy experiment above provides a controlled baseline to assess how instrumental resolution modulates bar recoverability. Building on the resolution-dependent trends identified in the borderline case, we restrict the sample-wide analysis to VIS mock images. Among the Euclid-like products considered, VIS is the only configuration that consistently preserves the key bar signatures—a clear ellipticity peak and a near-constant PA across the bar—enabling robust detections by Zoobot, Fourier $A_{2,\mathrm{mock}}$, and ellipse fitting; by contrast, $Y_{E}$ and VIS–NISP composites suppress these signatures due to PSF blurring and coarser sampling. Focusing on VIS therefore maximizes bar recoverability and maintains a uniform instrumental setup across the sample. All counts and trends reported in Figs.~\ref{fig:fig7}–\ref{fig:fig8} are thus measured on VIS mocks.

To highlight complementarity, Fig.~\ref{fig:fig8} assigns each galaxy to the first method that flags it as barred, using a fixed priority order that reflects supervision level: Zoobot (minimal), Fourier $A_{2}$ (low), and ellipse fitting (high). Under this combined view, 107 out of 141 systems are recovered, with the remaining galaxies not detected by any method.

\section{Discussion}
\label{discussion}

We organize the discussion in two parts: (i) a borderline barred galaxy (ID 184179, $z=0.5$) used to expose method- and filter-dependent systematics (Sec.~\ref{disc_borderline}); and (ii) a VIS-only, sample-wide analysis at $z=0.5$ that quantifies detectability and its impact on the apparent bar fraction (Sec.~\ref{disc_full_sample}).

\subsection{The borderline bar as a benchmark for detectability}
\label{disc_borderline}

In this step, we analysed the impact of adding realistic observational effects on the detectability of bars in a TNG50 galaxy. Our radiative transfer simulations considered Euclid-specific parameters, including filter transmission curves, PSFs, and representative noise levels. We generated realistic mock images for both the \(I_{E}\) (VIS) and \(Y_{E}\) (NISP) filters and compared the performance of different bar detection methods against the idealised case represented by the stellar mass map directly extracted from the TNG50 simulation.

Regarding the machine learning approach, we employed the Zoobot model, analogous to that used in the morphological analysis of real galaxies in the Euclid Q1 data release \citep{Walmsley2025}. While internal tests confirmed that Zoobot is effective in identifying visually prominent, extended bars, for our case-study galaxy the model only detected the bar in the higher-resolution $I_{E}$ mock image. As expected, the bar was not detected in the lower-resolution $Y_{E}$ mock. More interestingly, the VIS-NISP composite RGB image also failed to reveal the bar. Despite previous studies suggesting that VIS resolution dominates morphology classification in such RGB composites, our results indicate that the smoothing introduced by the inclusion of the lower-resolution $Y_{E}$ filter was sufficient to suppress the bar signature in this case.

For the ellipse fitting method, the stellar mass map served as an idealised benchmark. In the mocks, the bar was measurable in the $I_{E}$ filter but undetectable in $Y_{E}$ due to insufficient resolution. Similarly, the VIS-NISP composite image failed to recover the bar, as the rebinning and PSF convolution introduced during the RGB creation process smoothed out the isophotal signatures that were otherwise visible in the $I_{E}$ image alone.

When using the Fourier decomposition method, the borderline galaxy is classified as barred in the VIS $I_{E}$ mock, with $A_{2,\max}\!\geq\!0.2$ and a peak that is slightly shifted to larger radius relative to the stellar mass map (Fig.~\ref{fig:fig6}), whereas in $Y_{E}$ and in the VIS–NISP composite the $m\!=\!2$ amplitude falls below the detection threshold and no bar is recovered. This mirrors the behaviour seen with ellipse fitting—both methods succeed in VIS and fail in the other mocks.

This step of our paper was focused on one galaxy. To put its size in context, we briefly discuss the distributions of bar length in TNG50. Bar lengths are presented in figure 4 of \citet{Rosas-Guevara2022} as a function of stellar mass for different redshifts. The following quoted values refer to stellar masses in the range $10^{10-11.2}\,{\rm M_\odot}$, but there is no strong dependence on mass within this range. The bars at $z=0.5$ are found by \citet{Rosas-Guevara2022} to have median lengths close to 2 kpc, hardly reaching 3 kpc. At $z=1$ the median bar lengths are below 2 kpc. At $z=2-4$, the median is close to 1 kpc.

The galaxy ID184179 we chose for our present analysis has a bar length of 2.1 kpc at $z=0.5$, meaning that it is a fairly representative bar size of its redshift. And yet, this particular bar proved to be borderline undetectable, depending on the filters and methods. This results strongly suggests that more than half of the TNG50 bars at $z=0.5$ would likewise not be seen by Euclid. Therefore, the predicted TNG50 bar fraction of 44 per cent at $z=0.5$ should be expected to drop by a factor greater than two, once the same analysis is repeated for the full sample. If this estimate holds, then the corrected bar fraction of roughly 22 per cent would naturally reconcile the TNG50 prediction with the Euclid measured bar fraction at that redshift.

A similar behaviour should be expected once the corresponding corrected fractions are calculated for the $z=1$ TNG50  galaxies. For the even higher redshift range $z=2-4$, it is worth noticing that there are virtually no bars larger than 2 kpc in figure 3 of \citet{Rosas-Guevara2022}. Given the Euclid pixels sizes at that redshift range, it seems probable that the bar fraction will drop to nearly zero at those high redshifts.

Such expected corrections, of which the galaxy presented in this paper is one example, are similar in concept to applying a constant cut of minimum bar length throughout the entire sample. In other words, if the simulation analysis is limited to long bars ($r>2$ kpc), then the expected bar fractions will naturally decrease, as the undetectable short bars would not artificially inflate the simulated bar fraction. A further step would be to apply a non-uniform cut, i.e.~a redshift-dependent minimum threshold of bar length to be considered, informed by the PSF pixel size at each redshift (such as approximately twice the PSF FWHM; e.g. \citealt{Aguerri2009, Liang2024}). This would prevent the artificial decrease of the bar fraction close to $z=0$, where even the short bars are potentially detectable.

\subsection{Full sample bar detectability}
\label{disc_full_sample}

\begin{table}
\centering
\caption{Bar detections on VIS mocks at $z=0.5$ (TNG50). The combined row uses the supervision-based priority Zoobot $\rightarrow$ $A_{2}$ $\rightarrow$ ellipses; each galaxy is counted by the first method that flags it as barred.}
\label{tab:vis_recovery}
\begin{tabular}{l l c}
\toprule
Method & Criterion / Input & Recovered \\
\midrule
Zoobot & $p_{\mathrm{bar}}\ge 0.5$ & $31/141$ \\
Fourier $A_{2}$ & $A_{2,\max}\ge 0.2$ & $80/141$ \\
Ellipse fitting & $\epsilon$ peak + PA plateau & $67/141$ \\
Combined (priority) & Zoobot $\rightarrow A_{2}\rightarrow$ ellipses & $107/141$ \\
\bottomrule
\end{tabular}
\end{table}

Guided by the borderline case, we evaluated the full $z=0.5$ sample using VIS mock images only, since VIS is the only Euclid-like configuration that consistently preserves the key bar diagnostics in all methods (Zoobot detectability, $A_{2,\mathrm{mock}} \geq 0.2$ and a clear ellipticity peak and near-constant PA across the bar) in our case study.

The three methods differ markedly in their supervision requirements. Zoobot is designed to automatically classify large batches of images with minimal human intervention. By contrast, ellipse fitting (IRAF/Ellipse\footnote{IRAF is distributed by the National Optical Astronomy Observatory, which is operated by the Association of Universities for Research in Astronomy (AURA) under a cooperative agreement with the National Science Foundation.} or photutils.isophote from astropy) typically demands user oversight---e.g. centering, masking star-forming regions and judging convergence---and, in a non-negligible subset of galaxies, no stable isophotal solution is obtained (our “No fit” category) because the light distribution departs from ellipses. Fourier $A_{2}$ profiles, in turn, can usually be generated with little supervision once the geometric setup is fixed, and are comparatively robust to moderate morphological complexity.

Bar detections on the VIS mocks at $z=0.5$ yield recovered fractions of 31/141 with Zoobot, 79/141 with the Fourier $A_2$ method, and 66/141 from ellipse fitting, with the combined priority scheme reaching 107/141 (Table \ref{tab:vis_recovery}). A simple approximation based on the hypothesis that bars shorter than $2\times \mathrm{FWHM}$ (VIS) are undetectable would predict 67/141 detections, which happens to fall near the average of the values obtained with the different methods. However, this agreement should not be viewed as definitive, since bar size alone does not fully determine detectability; as shown in Fig.~\ref{fig:fig7}, some bars longer than this limit remain undetected, while others shorter than it are successfully recovered. Bar strength also plays a decisive role in detectability: weak bars can fall below the detection threshold even when they are formally longer than $2\times \mathrm{FWHM}$. Therefore, while bar size relative to the instrumental resolution sets a necessary condition, it is the combined distribution of bar length and strength that ultimately shapes the effective bar fractions obtained with different detection methods.

The Euclid Collaboration adoption of Zoobot is fundamentally driven by scale: even the Q1 \citep{EuclidCollab2025} release already contains $\sim$26 million detections over $63.1~\mathrm{deg}^2$, with later releases expected to grow toward $>10^9$ galaxies—making fast, uniform, ML-based classifications essential. In practice, probability thresholds, the Galaxy Zoo–based training domain, and instrument-specific inputs (passbands, PSF, sampling) tend to favour precision over completeness, so short/weak or low-contrast bars at VIS resolution are preferentially missed. A practical step is simply to complement Zoobot with Fourier $A_{2}$ and ellipse-based checks; even with light supervision, these diagnostics can recover barred systems that a single classifier may overlook and thereby raise the overall count.

Bar fraction estimates taken directly from stellar mass maps in cosmological simulations are over-optimistic because they ignore the instrument detectability function. In mass maps, short and/or weak bars can be “visible by mass resolution,” whereas in real data the PSF convolution, noise, spatial sampling, and bandpass differences smear the diagnostics (ellipticity peak, PA plateau, and the $A_{2}$ Fourier amplitude), pushing many systems below detection thresholds. This yields an upward bias when such fractions are compared naively to observations. Consequently, theory–observation comparisons should be based on radiative-transfer mocks with realistic instrumental conditions and apply the same classification criteria (e.g. Zoobot/$A_{2}$/ellipse fitting) to estimate the apparent fraction and, ideally, to model selection effects in order to infer the intrinsic fraction.

In face of these considerations, the method proposed in this paper has the advantage of aiming to perform quantitative measurements on individual galaxies. Rather than applying a fixed cut, or even a given cut at each $z$, this approach would provide a more reliable statistic by taking advantage of the natural diversity of morphologies and formation histories in the TNG50 galaxies. The value of the intrinsic bar length alone might not be the sole factor determining detectability. For example, there might be various morphological peculiarities that also influence bar detectability, such as spiral arms, rings, star-forming clumps---all of them also potentially waveband-dependent. Indeed, \citet{Menendez-Delmestre2024} showed that in observations bar length can be waveband-dependent, and \citet{goncalves2025} found that this behaviour is also present in TNG50 mock images. For these reasons, direct measurement of the individual mock images of the entire TNG50 barred sample is desirable. Furthermore, the procedures described here ensure that the observational effects are taken into account realistically at each filter.

% caveat
Apparent discrepancies between observational results and the expectations of cosmological simulations are often understood as providing indications for the refinement of theoretical models. Even if future analyses show that the evolution of TNG50 bar fractions can be consistenly reconciled by an unbiased comparision to observtions, this does not necessarily rule out other concerns such as the rate of stellar mass build-up in simulations or the growth and sizes of stellar discs at early times \citep[e.g.][]{Costantin2023}.

\subsection{Zoobot mock classification: caveats and perspectives}
\label{zoboot_perspectives}

The first caveat of this subsection is that, by construction, Zoobot is trained to classify real galaxies \citep{Walmsley2023}. We have aimed to achieve a high level of synthetic realism by reproducing the PSF, pixel size, noise, and filters through radiative transfer, which brings the mocks much closer to real observations than simple mass maps or magnitudes derived directly from the particle distribution. Nevertheless, our objects are not real galaxies. The key point is that the bar fractions obtained in our sample should not be directly extrapolated to the use of Zoobot in real observational data, including, in particular, the fractions reported for Euclid. However, even though the absolute values of the fractions are affected by this limitation, the fact that other methods show greater sensitivity to weak and short bars suggests that this behaviour should also manifest in real observations, albeit with different absolute values, indicating that a considerable number of bars may be overlooked by the current Euclid Q1 classification approach.

The second caveat is that we used the public Zoobot model, trained on earlier versions of Galaxy Zoo \citep[e.g.][]{2022Walmsley, walmsley2023_DESI}. In contrast, the bar fraction reported by Euclid Q1 \citep{EuclidCollab2025} was derived using a version of Zoobot that underwent fine-tuning with Euclidized \citep{Aussel2024} images, which are not currently available outside the collaboration. Therefore, it is plausible that the number of barred galaxies detected in our work would have been higher if a similar fine-tuning process had been applied.

Still in the context of fine-tuning, we believe that the presence of galaxies above the instrumental limit that are not classified as barred indicates the need for retraining. In particular, it would be possible to retrain the model using these galaxies as labels, thereby increasing its sensitivity to objects that, due to instrumental limitations, are incorrectly classified as unbarred. As a future perspective, we intend to explore the use of galaxies that lie below the detectability limit but can be inferred to be barred based on the numerical simulations, as a set of labels for fine-tuning and retraining not only Zoobot but also other morphological classification models with similar approaches, with the objective of increasing the number of detections not only in mocks but also in real galaxies.

\subsection{Sample selection of barred galaxies in TNG50}
\label{caveats_sample}

As discussed earlier, our analysis is restricted to galaxies with $M_\star \geq 10^{10}~{\rm M_\odot}$, following the mass selection adopted by \citet{Rosas-Guevara2022}. The results presented by \citet{2022Zana} provide an important reference for interpreting our measurements, particularly regarding the role of sample selection and stellar-mass range in shaping the inferred evolution of the bar fraction. As shown in that paper, variations in the adopted criteria, such as excluding the most massive systems or extending the sample to lower stellar masses, can lead to different redshift trends. In particular, the inclusion of a large population of low-mass galaxies, often hosting small and weak bars, contributes to the sensitivity of the measured bar fraction to the adopted selection.

In the context of our analysis, however, we do not expect that extending the sample to lower stellar masses would significantly alter our results. As discussed by \citet{2022Zana}, galaxies with $M_\star < 10^{10}~{\rm M_\odot}$ typically exhibit weak bar signatures, with $A_{2,\max} < 0.2$. These values are derived from noise-free stellar-mass distributions, where non-axisymmetric features are measured directly from the particle distribution using \textsc{MORDOR}. While this approach is fully self-consistent within that framework, it differs fundamentally from analyses based on projected flux.

When relying on mock observations designed to mimic real datasets, such weak bar signatures are expected to be strongly suppressed once synthetic noise, PSF convolution, and projection-induced smoothing are included. As a result, the detectability of these low-amplitude $A_2$ features is substantially reduced in our mock-based analysis. In addition, a fraction of these low-mass systems are resolved with too few stellar particles at the mass resolution of \textsc{TNG50}, which limits the construction of reliable radiative-transfer mocks and prevents robust morphological measurements. These considerations justify our focus on galaxies with $M_\star \geq 10^{10}~{\rm M_\odot}$.\\

\subsection{Impact of reclassification on bar ellipticity and length}
\label{ellipticity_lengths}

Beyond detectability, reclassification also impacts the interpretation of measured bar structural parameters. \citet{goncalves2025} presented measurements of bar ellipticities and lengths in TNG50 based on realistic radiative-transfer mock images generated with \textsc{SKIRT} \citep{Baes2015}. That paper demonstrated that both quantities depend on the adopted wavelength, with bars in star-forming galaxies of TNG50 becoming progressively longer and more elliptical when measured in bluer filters. This behaviour reflects the increasing contribution of young stellar populations and highlights the limitations of comparing structural properties measured at different wavelengths without accounting for observational effects. When compared to measurements based on stellar-mass maps, same bars identified in realistic mock images can appear up to $\sim$ 20  per cent longer and more elliptical.

Although our focus is on bar detectability, we adopt noise-free stellar-mass measurements of bar length and $A_2$ from \citet{Rosas-Guevara2022} as reference values for characterisation. We note, however, that bar properties depend on the measurement technique and differ when realistic mock observations are considered. If only stellar mass maps are used, bars are typically longer and more elliptical, while the inclusion of synthetic observational effects, such as PSF convolution and noise, suppresses weak bars and smooths the strongest ones. This leads to an overall reduction in the detected bar fraction. However, on a galaxy-by-galaxy basis, bars that remain detectable after these effects are applied can appear longer and more elliptical than their counterparts measured from stellar-mass maps. This behaviour is consistent with the wavelength-dependent trends reported by \citet{goncalves2025} for TNG50 when comparing mock-based measurements, which include the effects of different stellar populations distributed along the bar, with those derived from stellar-mass maps.

A natural extension of our work, currently in development, is to quantify the impact of bar reclassification on the mean bar ellipticity and length over a broader redshift range, rather than focusing on a single epoch at $z=0.5$. In this context, future analyses will explore mock observations tailored not only to Euclid-like conditions but also to JWST at higher redshifts. This approach therefore enables a comparison of bar lengths and ellipticities at high redshift that is more closely aligned with the observational context. To date, such comparisons have been carried out predominantly using noise-free stellar-mass maps, as in the study of \citet{Guo2025}, which adopts the bar lengths from \citet{Rosas-Guevara2022} and compares them with JWST/NIRCam observations. However, in the present work our primary focus remains on detectability and classification completeness rather than on detailed structural recalibration.

\section{Conclusions}
\label{conclusions}

In this paper, we addressed the question of whether TNG50 barred galaxies would be detectable by Euclid at high redshift. As a case study, we focused on one representative simulated galaxy at $z=0.5$. We produced realistic mock images in Euclid filters and evaluated the bar detectability using three methods. Regarding this example galaxy, our results may be summarised as follows:

\begin{table}
\centering
\caption{Apparent bar fractions at $z=0.5$ with a fixed denominator $N=318$ galaxies from \citet{Rosas-Guevara2022}. Non-detections are counted as unbarred in our bar fractions.}
\label{tab:apparent_barfractions_318}
\begin{tabular}{lc}
\toprule
Method & Barred $N$ (percentage) \\
\midrule
\multicolumn{2}{l}{\textit{Baseline (mass maps) from \citet{Rosas-Guevara2022}}} \\
Mass maps & 141 (44 per cent) \\
\midrule
\multicolumn{2}{l}{\textit{This work (VIS mocks; non-detections $\rightarrow$ unbarred)}} \\
Zoobot & 31 (10 per cent) \\
Fourier $A_{2,\mathrm{mock}}$ & 80 (25 per cent) \\
Ellipse fitting & 67 (21 per cent) \\
Combined (Zoobot $\rightarrow A_{2} \rightarrow$ ellipses) & 107 (34 per cent) \\
\bottomrule
\end{tabular}
\end{table}

\begin{enumerate}

\item The Zoobot machine learning classifier only recovers the bar in the VIS $I_{E}$ band, but not in the NISP $Y_{E}$ band, nor in other combinations of filters that include $Y_{E}$.

\item Ellipse–fitting preserves the characteristic bar signatures (PA plateau and a clear ellipticity peak) only in VIS \(I_{E}\), with the peak ellipticity even higher than in the stellar mass map; in \(Y_{E}\) and in VIS–NISP composites the ellipticity remains low and lacks a bar–related maximum.

\item In the Fourier decompositions, the bar is recovered only in VIS \(I_{E}\), with \(A_{2,\max} > 0.2\) and a peak shifted to slightly larger radius than in the mass map; in \(Y_{E}\) and in VIS–NISP composites the $(A_{2,\max})$ amplitude falls below the detection threshold.

\item The conclusion is that the bar detectability of a typical galaxy at $z=0.5$ is dependent on the instrument and on the method. This particular 2.1\,kpc bar was borderline identifiable in the high-resolution VIS $I_{E}$. However, most bars at $z>0.5$ are shorter and weaker than this example.

\end{enumerate}

These results have relevant implications for interpreting the bar fraction in TNG50. It is well established that TNG50 presents a bar fraction at $z=0.5$ that exceeds observational estimates, with a higher bar fraction at $z=0.5$ than at $z=0$, in disagreement with trends observed in real surveys such as Euclid. Our proof-of-concept demonstrates that applying realistic observational effects can render some bars undetectable, emphasizing that direct comparisons between simulated mass distributions and observed bar fractions---without accounting for detectability biases---may lead to misleading conclusions. Notably, ID184179 occupies an intermediate position in the bar parameter space at this redshift, with many galaxies in TNG50 hosting even smaller and weaker bars. Given this, it is likely that a significant fraction of bars in the simulation would go undetected under Euclid-like conditions, effectively lowering the observable bar fraction.

Extending beyond the borderline case, we evaluate the full $z=0.5$ sample on VIS-only mock images to maximise bar recoverability and maintain a uniform instrumental setup. The same three diagnostics—Zoobot, Fourier $A_{2}$, and ellipse fitting—are applied consistently and summarised in the bar radius–strength plane (Figs.~\ref{fig:fig7}–\ref{fig:fig8}), using a supervision-based priority (Zoobot $\rightarrow$ $A_{2}$ $\rightarrow$ ellipses) for the combined view, with per-method recoveries listed in Table~\ref{tab:vis_recovery}. Regarding the full $z=0.5$ sample, our results may be summarised as follows:

\begin{enumerate}

\item Missed detections for all methods cluster at low strength and small size, i.e.\ $A_{2}\!\lesssim\!0.4$ and $L_{\mathrm{bar}}\!\lesssim\!2\,\mathrm{kpc}$. Zoobot additionally misses a tail of long/strong systems, whereas ellipse fitting also fails whenever no stable isophotal solution is obtained (``No fit'' cases).

\item In our simulated RT mock sample complementing Zoobot with Fourier $A_{2}$ and ellipse fitting increases the number of barred galaxies recovered, but many short/weak systems ($A_{2}\!\lesssim\!0.4$, $L_{\mathrm{bar}}\!\lesssim\!2$\,kpc) remain below detectability even under VIS conditions.

\item Thus, the apparent bar fraction spans $12$--$33$ per cent depending on method (Zoobot, Fourier $A_{2,\mathrm{mock}}$, ellipse fitting, and the combined priority scheme; see Table~\ref{tab:apparent_barfractions_318}).

\item For Zoobot, the lower sensitivity to short/weak bars is expected to also occur in real observations, implying that current bar fractions (e.g.\ Euclid Q1) may be underestimated and warrant further investigation.

\end{enumerate}

Fine-tuning the model with Euclid-like data has the potential to mitigate this underestimation, with RT mocks providing a valuable source of labels for retraining Zoobot and similar morphological classifiers. As future work, we plan to extend this analysis to all TNG50 subhalos at lower and higher redshifts, aiming to derive a more realistic, observationally-aware bar fraction that accounts for the full range of instrumental and observational effects. This combined approach will enable more reliable comparisons between simulations and current or future galaxy surveys, including Euclid, JWST, and others.

\section*{Acknowledgements}

GG acknowledges support from  \textit{Coordenac\c c\~ao de Aperfei\c coamento de Pessoal de N\'ivel Superior} -- Brasil (CAPES) -- Finance Code 001, and from the Brazilian agency \textit{Conselho Nacional de Desenvolvimento Cient\'ifico e Tecnol\'ogico} (CNPq). GG thanks the IllustrisTNG collaboration for generously providing access to TNG50 simulation data and computational resources via the online JupyterLab workspace. RM acknowledges support from CNPq through grants 406908/2018-4 and 307205/2021-5, and from \textit{Funda\c c\~ao de Apoio \`a Ci\^encia, Tecnologia e Inova\c c\~ao do Paran\'a} through grant 18.148.096-3 -- NAPI \textit{Fen\^omenos Extremos do Universo}.

\section*{Data availability}
The data underlying this article will be shared on reasonable request to the corresponding author.

\bibliographystyle{mnras.bst}
\bibliography{paper.bib}

\bsp
\label{lastpage}

\end{document}